\documentclass[referee, pdflatex, sn-nature]{sn-jnl-modified}

\usepackage{graphicx}%
\usepackage{multirow}%
\usepackage{amsmath,amssymb,amsfonts}%
\usepackage{amsthm}%
\usepackage{mathrsfs}%
\usepackage[title]{appendix}%
\usepackage{xcolor}%
\usepackage{textcomp}%
\usepackage{manyfoot}%
\usepackage{booktabs}%
\usepackage{algorithm}%
\usepackage{algorithmicx}%
\usepackage{algpseudocode}%
\usepackage{listings}%

\begin{document}

\title[Article Title]{Investigating the relationship between the Weyl semimetal phase and the three-dimensional quantum Hall phase in ZrTe$_5$}


\author[1]{\fnm{Jiahao} \sur{Chen}}

\author[2,3]{\fnm{Yu} \sur{Cao}}

\author[2]{\fnm{Hong} \sur{Du}}

\author[1]{\fnm{Yuanze} \sur{Li}}

\author[2,3]{\fnm{Ruidan} \sur{Zhong}}

\author*[1,4]{\fnm{Tian} \sur{Liang}}\email{tliang@mail.tsinghua.edu.cn}

\affil[1]{\orgdiv{State Key Laboratory of Low Dimensional Quantum Physics, Department of Physics}, \orgname{Tsinghua University}, \postcode{100084}, \orgaddress{\city{Beijing}, \country{China}}}

\affil[2]{\orgdiv{Tsung-Dao Lee Institute}, \orgname{Shanghai Jiao Tong University}, \postcode{201210}, \orgaddress{\city{Shanghai}, \country{China}}}

\affil[3]{\orgdiv{School of Physics and Astronomy}, \orgname{Shanghai Jiao Tong University}, \postcode{200240}, \orgaddress{\city{Shanghai}, \country{China}}}

\affil[4]{\orgname{Frontier Science Center for Quantum Information},  \postcode{100084}, 
\orgaddress{\city{Beijing},
\country{China}}}


\abstract{
The material ZrTe$_5$ exhibits distinct topological phases, including a Weyl semimetal phase, characterized by a chiral anomaly and in-plane Hall effect, and a three-dimensional quantum Hall phase. The relationship between these phases remains poorly understood. This work systematically explores their connection in ZrTe$_5$ through rotatable, pressure-dependent measurements. At ambient pressure, both phases are observed; the WSM phase requires strong electronic polarization, while the 3D QH phase appears when the characteristic resistivity peak temperature \textit{$\text{\rmfamily T}_{\text{\rmfamily p}}$} is approximately 90 K. Under applied pressure, the polarization diminishes, weakening the WSM phase and its associated nontrivial Hall signals. Concurrently, \textit{$\text{\rmfamily T}_{\text{\rmfamily p}}$} rises dramatically from 2 K at ambient pressure to 70 K at 2.2 GPa, approaching the expected regime for the 3D QH phase. These findings clarify the conditions underlying the WSM and 3D QH phases and suggest that exploring the 3D QH phase at even higher pressures is a promising direction for future research.

}


\maketitle

\section{Introduction}
Topological phases of matter have become a central topic of research within condensed matter physics over the past twenty years, attracting widespread interest due to their unique and often exotic properties\cite{Kane2010,SCZhang2011}. In recent years, zirconium pentatelluride (ZrTe$_5$) has emerged as a particularly intriguing material within the broader family of topological materials. Its significance stems from the rich variety of physical phenomena it exhibits, many of which are closely linked to its non-trivial topological nature. Notably, ZrTe$_5$ has been observed to display the chiral anomaly\cite{Qi2016,Liang2018}, a quantum phenomenon manifesting as a negative longitudinal magnetoresistance\cite{OngReview}. Furthermore, the material exhibits an in-plane Hall effect\cite{Liang2018}, which originates from the Berry curvature in the Weyl semimetal (WSM) phase, distinct from ordinary Hall effects arising from Lorentz force which requires perpendicular applied magnetic fields. In addition to these phenomena, ZrTe$_5$ also shows signatures of a three-dimensional quantum Hall (3D QH) effect\cite{Zhang2019,Irvine}, where 2D QH\cite{Klitzing} layers are stacked along a particular crystallographic direction in real space\cite{Halperin}. These multifaceted physical responses have propelled ZrTe$_5$ to the forefront of modern condensed matter research, motivating extensive experimental and theoretical investigations aimed at unraveling its underlying topological nature\cite{X.J.Zhou,Ando,Gooth,ZhangLiYuan2}.

Structurally, ZrTe$_5$ is predicted to crystallize in a specific orthorhombic lattice, belonging to the \textit{Cmcm} space group $(D_{2h}^{17})$, as depicted schematically in the inset of Fig. 1a. This crystal structure features lattice constants that place ZrTe$_5$ very close to the phase boundary separating weak topological insulator (WTI) and strong topological insulator (STI) regimes\cite{XiDaiPRX}. However, actual experimental observations have revealed that rather than exhibiting purely WTI or STI behavior, ZrTe$_5$ appears to stabilize in phases that are better described as WSM\cite{Qi2016,Liang2018} or 3D QH phases\cite{Zhang2019}. The emergence of these phases points to the presence of additional order parameters within the system, which break certain symmetries and enable new topological behaviors beyond those expected from the nominal crystal structure alone.

Focusing on the WSM phase, the electronic structure is characterized by pairs of Weyl nodes—points in momentum space where conduction and valence bands touch linearly, acting as sources and sinks of Berry curvature\cite{Murakami_2007,Murakami2}. In the right panel of Fig. 1c, it is illustrated how two-dimensional slices of momentum space, taken perpendicular to the axis connecting these Weyl nodes, form 2D QH states. Each such 2D plane can be viewed as a 2D QH system with a non-zero Chern number. One of the hallmarks of this phase is the formation of Fermi arcs\cite{Arc1,Arc2} on the surfaces of the material. These arcs originate from the one-dimensional chiral edge states inherent to the 2D QH states in momentum space and appear on every surface of the three-dimensional crystal.

In contrast, the 3D QH phase is conceptually different. Here, the system can be thought of as a stack of 2D QH layers arranged along a particular crystallographic direction in real space\cite{Halperin}. Unlike the WSM phase where the edge states appear on all surfaces due to the existence of Weyl nodes, the chiral edge states in the 3D QH phase manifest only on specific surfaces corresponding to the edges of these stacked 2D layers, having a sharp difference from those of the WSM phase.

Even though information about edge states can, in principle, distinguish the WSM phase from the 3D QH phase, it is practically very difficult to obtain such information experimentally. For the WSM phase, the typical chemical potential energy required to resolve the Fermi arcs experimentally is very low, typically on the order of a few tens of meV or below\cite{Ando,Ong(Xiong)Science}. This makes the direct experimental detection of the Fermi arcs extremely challenging. In contrast, for the 3D QH phase, the edge states exist on the side surfaces of the sample and are experimentally inaccessible via spectroscopic measurements. Therefore, distinct experimental features other than edge states are needed to differentiate the WSM phase from the 3D QH phase. 

An important experimental observation is that the WSM phase supports the presence of in-plane Hall signals, which are measurable in transport experiments. On the other hand, the 3D QH phase does not exhibit such in-plane Hall responses. This fundamental difference implies that these two phases are mutually exclusive and cannot coexist simultaneously within the same sample. In the case of actual ZrTe$_5$ crystals, as shown in Figs. 1c, the evolution of polarization and other order parameters facilitates a transition between the WSM and 3D QH phases. This tunability offers a unique platform to study the interplay between competing topological phases and investigate the mechanisms governing their stability and transitions, which will be elaborated upon in the following sections.

\section{Results}

\subsection{3D QH vs. Weyl Semimetal}
Before discussing the angular and pressure-dependent responses of ZrTe$_5$, we examine at ambient pressure two samples—ZT412 and ZT015—that highlight the distinctive properties of the 3D QH and the WSM phases, respectively.

As shown in Fig. 2b, under a magnetic field aligned along the b-axis (B$\parallel$b-axis, $\mathrm{\theta = 90^\circ}$, out-of-plane geometry), the ZT412$\#$2 sample exhibits a Hall resistivity ($\rho_{yx}$) featuring a kink near B ($\theta = 90^\circ$) $\approx$ 1.3T, where it enters the quantum limit. The value of $\rho_{yx}$($\theta$ = 90°) is approximately 7.5 $m\Omega$  $cm$. When the magnetic field is tilted from the out-of-plane direction (b-axis) toward the in-plane direction (c-axis), both the kink structure and the resistivity value remain nearly constant at around 7.5 $m\Omega$  $cm$, with the effective field scaling as B ($\theta$) $\approx$ B ($\theta = 90^\circ$)/$\sin\theta$. This strongly indicates the onset of the 3D QH phase in the ZT412$\#$2 sample. This conclusion is further supported by data from ZT412$\#$1 (Fig. 2c), which was measured on the same ZrTe$_5$ sample (ZT412$\#$2) approximately two years prior. Notably, the transition temperature \textit{$\text{\rmfamily T}_{\text{\rmfamily p}}$} is around 90 K, close to the reported value of $\sim$95K for the 3D QH phase in the literature\cite{Zhang2019}, suggesting that \textit{$\text{\rmfamily T}_{\text{\rmfamily p}}$} serves as an indicator of the underlying order parameter responsible for inducing the 3D QH phase in ZrTe$_5$.

In contrast, the ZT015 sample displays prominent in-plane Hall signals ($\rho_{yx}$ $\approx$ 13 $m\Omega$) as shown in Fig. 2e, indicating that the system resides in the WSM phase. This behavior sharply contrasts with the 3D QH phase, where no in-plane Hall signals are observed. The WSM phase and its associated in-plane Hall effect are closely linked to a significant polarization developing in the system, quantified as P $\propto$ $\gamma'/\chi$$=$$ 2AR_{2\omega} / (R_0BI_0\chi)$\cite{Ando,Rikken}, reaching a value of approximately $2.7\times10^{-9}T^{1}A^{-1}m^{2}$, as shown in Fig. 2f. The polarization breaks the inversion symmetry and when it becomes sufficiently large, the WSM phase emerges in ZrTe$_5$. Moreover, when the polarization has components along the c-axis (in-plane direction), as in the ZT015 sample, it breaks mirror symmetry relative to the ab-plane, enabling the appearance of in-plane Hall signals generated by the Berry curvature originating from the Weyl nodes\cite{Liang2021PbSnTe,LiangBookChapter2,HuangPRL}.

\subsection{Pressure Dependence}
As described above, at ambient pressure, both the 3D QH phase and the WSM phase are observed. In the WSM phase, strong polarization develops, resulting in pronounced in-plane Hall signals originating from the Berry curvature generated by the Weyl nodes. Conversely, in the 3D QH phase, polarization is suppressed (see supplemental information Fig. S3f for details), and an additional order parameter associated with the characteristic temperature \textit{$\text{\rmfamily T}_{\text{\rmfamily p}}$} emerges, causing a distinctive kink in the Hall resistivity.

Since the strength of polarization is expected to vary under applied pressure, it is a natural step to investigate how the ZrTe$_5$ sample evolves when subjected to different pressure conditions. To this end, we employed a rotatable pressure cell, as depicted in Fig. 1b, and conducted systematic angular- and pressure-dependent transport measurements on sample ZT006. The results are presented in Fig. 3 and in Supplemental Figs. S4, S5, and S6. Two types of pressure cells were used for these measurements: the rotatable pressure cell (Almax pressure cell), which covers pressures up to approximately 1 GPa, and a second, non-rotatable pressure cell (Quantum Design pressure cell) capable of reaching pressures as high as 2.5 GPa. Since the main features are adequately captured within the pressure range up to 1 GPa, the description in the main text focuses primarily on this range. For data and analysis at higher pressures, refer to the supplemental materials.

The Hall resistivity, $\rho_{yx}$, exhibits anomalous behavior across all measured angles, including the in-plane direction (0 degrees, aligned with the c-axis), throughout the pressure range from 0 to 1 GPa. This anomaly reflects the influence of Berry curvature generated by the Weyl nodes characteristic of the WSM phase. Although $\rho_{yx}$ remains on the order of a few milliohms throughout the pressure range considered, the anomalous portion of the Hall conductivity, defined as $\sigma^A_{xy}=n\langle \Omega \rangle$ (where $n$ is the carrier density and $\langle \Omega \rangle$ is the average Berry curvature), decreases sharply due to an increase in longitudinal resistivity under pressure. Assuming that the carrier density $n$ remains approximately constant with pressure, the rapid drop in $\sigma^A_{xy}$ indicates a reduction in the average Berry curvature. This behavior is accurately captured in Figs. 3d and 3e, where the Hall conductivity $\sigma_{xy}$ saturates at high magnetic fields, isolating the anomalous component $\sigma^A_{xy}$. This saturation underscores the influence of Berry curvature arising from the redistribution of Weyl nodes under applied magnetic fields.

Further evidence supporting these observations comes from the evolution of polarization, which is proportional to the parameter $\gamma'$. As shown in Figs. 3g and 3h, polarization along both the out-of-plane (b-axis) and in-plane (c-axis) directions decreases dramatically as pressure increases. This reduction signals a weakening of the WSM phase. Correspondingly, the magnitude of the anomalous Hall conductivity $\sigma^A_{xy}$, reflecting the Berry curvature, also diminishes along both axes, as illustrated in Fig. 3f. This trend is further corroborated by the data in Fig. 3i, where the nonlinear Hall signals—proportional to the Berry curvature dipole density—also decline in tandem with the weakening polarization.

\section{Discussion and Conclusion}

The detailed physical properties observed in ZrTe$_5$ under varying pressure conditions allow us to construct a comprehensive phase diagram, as shown in Fig. 4. This diagram encapsulates a wealth of information regarding the interplay between pressure, polarization, and the corresponding electronic phases.

As external pressure is increased, the strength of electronic polarization steadily decreases. This decline directly weakens the WSM phase, which relies on robust polarization and the presence of Weyl nodes. Consequently, all physical quantities intimately connected to the WSM phase—including the Berry curvature, the Berry curvature dipole density, and anomalous transport signals—also become suppressed. These diminishing quantities collectively define the boundary of the WSM phase in the phase diagram, represented by dashed lines.

Simultaneously, an intriguing and important feature emerges in the behavior of the characteristic temperature \textit{$\text{\rmfamily T}_{\text{\rmfamily p}}$}, defined by the temperature where resistivity peaks. Unlike polarization, \textit{$\text{\rmfamily T}_{\text{\rmfamily p}}$} increases monotonically with pressure, growing from approximately 2 K at ambient pressure to nearly 70 K at the highest pressure studied (2.2 GPa). Here we note that another sample, ZT412, which exhibits the 3D QH phase at ambient pressure, shows a \textit{$\text{\rmfamily T}_{\text{\rmfamily p}}$} of about 90 K. This pressure-induced increase of \textit{$\text{\rmfamily T}_{\text{\rmfamily p}}$} in the ZT006 sample suggests that the system is gradually driven toward the 3D QH phase as pressure is raised.

It is crucial to emphasize that the 3D QH phase and the WSM phase are fundamentally incompatible; they cannot coexist within the same parameter space. Therefore, for the 3D QH phase to emerge, the WSM phase must first vanish.  In the ZT006 sample, this scenario is observed as pressure exceeds $\sim$2.2 GPa: the WSM phase is nearly vanquished, which coincides with a significant reduction in polarization and a severe weakening of anomalous Hall signals. This pressure marks a critical threshold where the electronic system transitions away from the WSM regime.

However, the mere disappearance of the WSM phase does not guarantee the formation of the 3D QH phase. The latter requires the development of an additional order parameter, distinct from polarization, that stabilizes this new electronic state. Based on both the experimental observations reported here and elsewhere\cite{Zhang2019} and prior theoretical work\cite{Halperin}, this additional order parameter is most likely a type of density wave order, which is closely linked to the characteristic temperature \textit{$\text{\rmfamily T}_{\text{\rmfamily p}}$}.

While our study systematically measures \textit{$\text{\rmfamily T}_{\text{\rmfamily p}}$} as a proxy for this additional order parameter, direct experimental detection and characterization of the order parameter itself remain outstanding challenges. Understanding the nature of this order parameter is a promising direction for future research. Such investigations could elucidate the precise mechanisms governing the emergence of the 3D QH phase and clarify the exact transition pathway connecting the 3D QH and the WSM phases, possibly mediated by pressure or strain.

In summary, the present study of ZrTe$_5$ reveals a rich relationship between the 3D QH phase and WSM phase. Increasing pressure suppresses polarization and the associated WSM phase, while simultaneously promoting the conditions necessary for the emergence of the 3D QH phase. The phase diagram constructed here serves as a roadmap for further experimental and theoretical explorations aimed at understanding and controlling topological phases in ZrTe$_5$ and related materials.

\section{Methods}

\subsection{Sample Growth}
High-quality ZrTe$_5$ single crystals were grown by the self-flux technique. High-purity zirconium (Zr) and tellurium (Te) were mixed in a molar ratio of 1:300 and sealed in an evacuated quartz tube. The mixture was first heated up to 900 ℃ over 3 hrs and held there for 48 hrs. Subsequently, it was rapidly cooled to 600 ℃, followed by slow cooling to 469 ℃ at a rate of 1℃/hr. Finally, after maintaining the isothermal stage at 469 ℃ for 2 days, rapid centrifugation (spin) was carried out. Needle-like single crystals with a typical length up to $\sim$5 mm, cross sectional area $>$ 0.1 mm * 0.1 mm, can be obtained from the residuals. The 3D QH sample ZT412$\#$2 used for transport measurements has dimensions of length 0.93 mm, width 0.16mm, and thickness 0.21mm. The WSM sample ZT015 used for transport measurements has dimensions of length 1.7 mm, width 0.15mm, and thickness 0.10mm.

\subsection{Transport Measurement}
Electrical connections are made using gold wires and silver paste. Both linear and non-linear electrical transport measurements are conducted using Quantum Design 14T Dynacool. For six-terminal linear transport measurements, longitudinal and transverse resistances are measured using a K6221 AC current source and SR830 lock-in amplifiers. For non-linear transport measurements, signals are monitored by the lock-in technique with harmonic number = 2, relative to the excitation current (f = 13.777 Hz).

\subsection{Pressure Cell}
Two types of pressure cells, Almax pressure cell and Quantum Design (QD) pressure cell are used. 
\\
1. Almax pressure cell is 360 degree rotatable pressure cell which can go to $\sim$1GPa. Almax pressure cell is mounted on the sample rotator to perform angle-dependent measurements. In-plane direction is accurately calibrated as shown in the supplemental information.
\\
2. QD pressure cell can go to $\sim$2.5GPa and is not rotatable. The determination of in-plane direction in QD pressure cell is made by eye and therefore not accurate.

\backmatter





\section{Acknowledgements}
T.L. and R.Z. acknowledge the support for the project by the National Key R$\&$D Program of China (No. 2021YFA1401600) fund. The work was also supported by the Ministry of Science and Technology of China under 2022YFA1402702, National Natural Science Foundation of China with Grants No. 12334008, and No. 12374148.

\section{Author contributions}
T.L. conceived and designed the project, provided overall direction, supervised the experiments, and coordinated collaborations among research groups. T.L. also proposed the initial idea of the rotatable pressure cell compatible with a low-temperature horizontal rotator. J.C. performed all pressure-dependent measurements with in-depth discussions with T.L. Y.C. H.D. and R.Z. grew the ZrTe$_5$ single crystals, which were subsequently characterized and screened by J.C. with support from Y.C. H.D. and R.Z. J.C. and T.L. conducted analyses of measurements and wrote the manuscript with input from all authors.Y. L. provided assistance in the LabVIEW data collection. All authors discussed the results and provided feedback on the manuscript.

\section{Competing interests}
The authors declare no competing interests.

\bigskip







\bibliography{MainPaper}
\begin{figure}[htbp]
  \centering
  \includegraphics[width=1\linewidth]{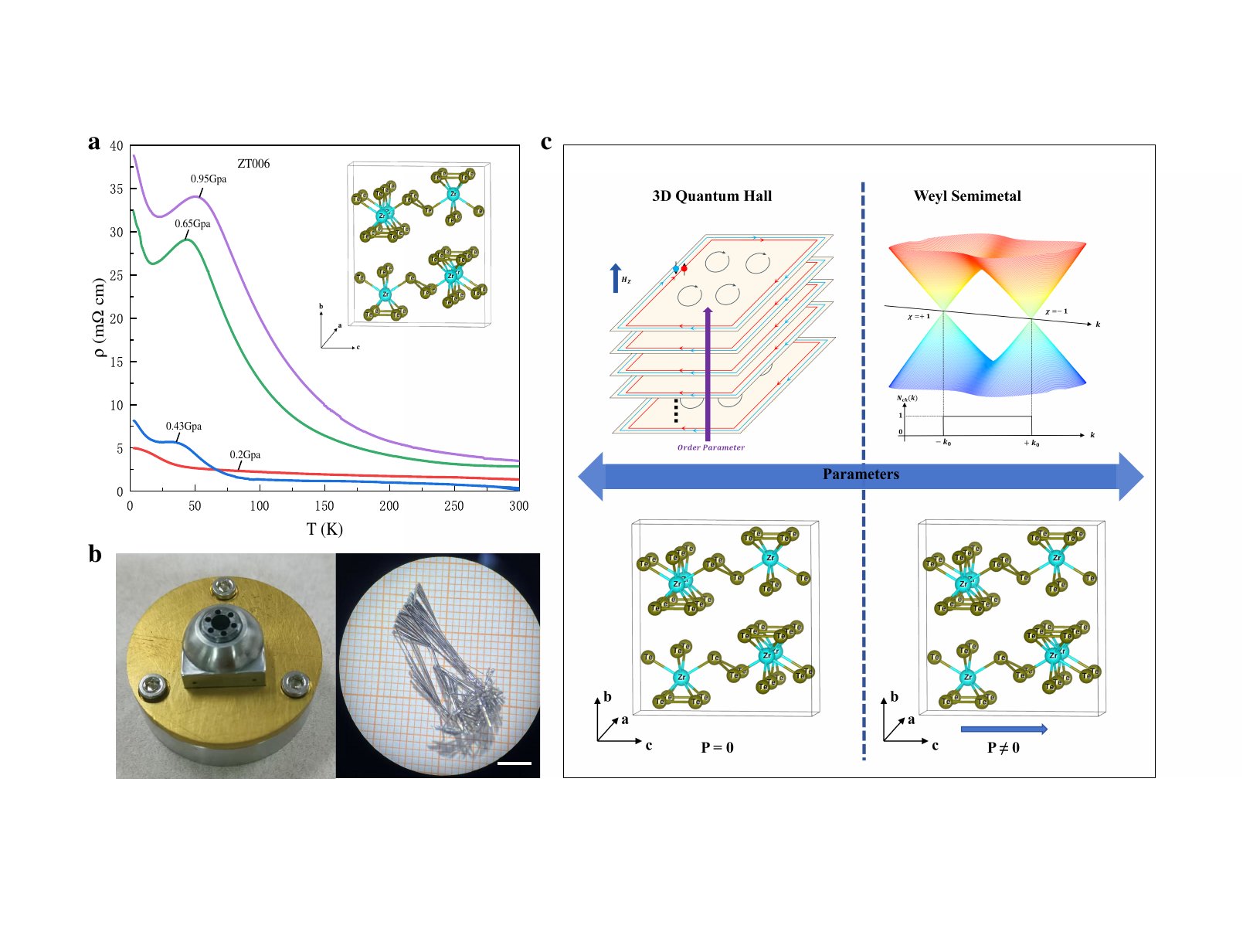}
  \caption{\textbf{Resistivity under different hydrostatic pressures.} 
  \textbf{a}, Resistivity ($\rho$) versus temperature (T) for ZrTe$_5$ under varying hydrostatic pressures (sample ZT006). The peak temperature ($\text{\rmfamily T}_{\text{\rmfamily p}}$) increases with pressure. The inset shows the crystal structure of ZrTe$_5$. \textbf{b}, \textit{Left panel}: A pressure cell designed for 360-degree rotation measurements at low temperatures and under applied magnetic fields. \textit{Right panel}: A microscope image of the ZrTe$_5$ single crystal. Scale bar is 5 mm. \textbf{c}, Schematic drawing of the Weyl semimetal (WSM) and 3D quantum Hall (3D QH) phases connected via order parameters. The 3D QH phase consists of stacked 2D QH layers along a specific crystallographic direction in real space, whereas the WSM phase features Weyl node pairs in momentum space, acting as Berry curvature sources/sinks. Each 2D momentum-space slice corresponds to a 2D QH system characterized by a Chern number. Crucially, the two phases are distinguished by the in-plane Hall effect: the WSM phase exhibits this effect (indicating existence of polarization), while the 3D QH phase does not. Tuning the order parameters drives transitions between these phases.
  }
  \label{fig:fig1}
\end{figure}

\begin{figure}[htbp]
  \centering
  \includegraphics[width=1\linewidth]{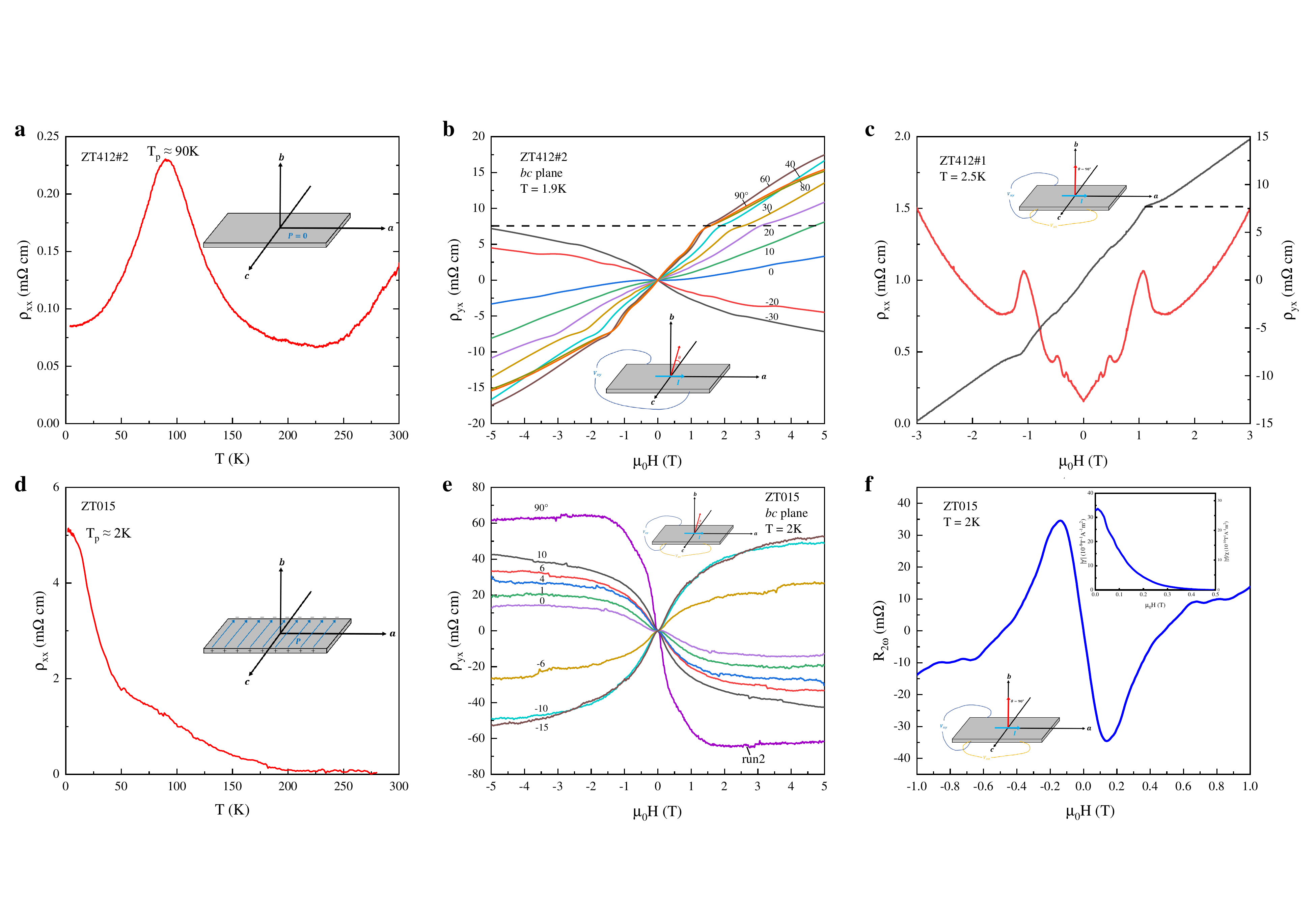}
  \caption{\textbf{3D QH effect (sample ZT412) and in-plane Hall effect (sample ZT015) at ambient pressure.} 
 Magnetic field angle ($\theta$) definitions are provided in the insets. \textbf{a}, $\rho$ vs. T for the 3D QH sample ZT412 (\textit{$\text{\rmfamily T}_{\text{\rmfamily p}}$} $\approx$ 90 K). \textbf{b}, Angle-dependent $\rho_{yx}$ vs. $\mu_0H$ for ZT412$\#$2 (field rotated in the bc plane). The near-constant resistivity ($\sim$7.5 $m\Omega$  $cm$, dashed line) confirms the 3D QH phase. \textbf{c}, $\rho_{yx}$ vs.$\mu_0H$ for ZT412$\#$1. The matching positions of the kink in Hall resistivity and the entrance into the quantum limit provide further confirmation of the 3D QH phase. \textbf{d}, $\rho$ vs. T for the in-plane Hall sample ZT015 (\textit{$\text{\rmfamily T}_{\text{\rmfamily p}}$} $\approx$ 2 K). \textit{Inset}: Polarization direction along the c-axis. \textbf{e}, Angle-dependent $\rho_{yx}$ vs. $\mu_0H$ for ZT015 (field rotated in the bc plane). A prominent in-plane Hall signal as large as $\rho_{yx}$ $\approx$ 13 $m\Omega$ signifies the WSM phase. \textbf{f}, Second-harmonic resistance ($R_{2\omega}$) vs. $\mu_0H$ (along the b-axis) for ZT015. Insets: Calculated $\left|\gamma'\right|$ and the ratio $\left|\gamma'\right|/\chi$ (representing polarization strength) vs. $\mu_0H$, showing the development of prominent electrical polarization along the c-axis (the in-plane direction).}\label{Fig. 2
  }
  \label{fig:fig2}
\end{figure}

\begin{figure}[htbp]
  \centering
  \includegraphics[width=1\linewidth]{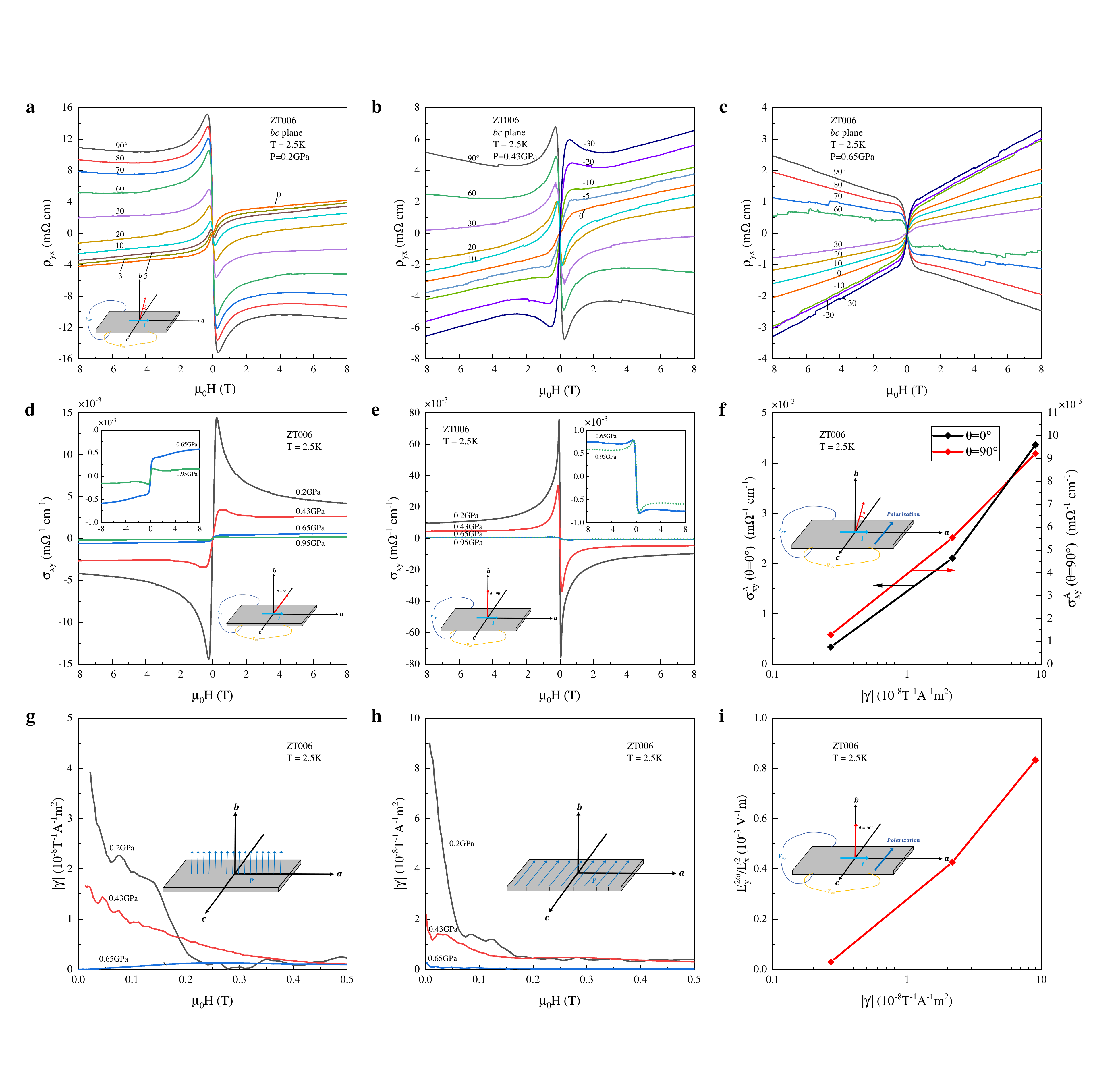}
  \caption{\textbf{Angle- and pressure-dependent measurements for sample ZT006 in a rotatable pressure cell.} 
 \textbf{a–c}, Angle-dependent Hall resistivity ($\rho_{yx}$) vs. magnetic field ($\mu_0H$) at 0.2 GPa, 0.43 GPa, and 0.65 GPa, respectively. A prominent in-plane Hall signal is observed at all pressures. \textbf{d}, \textbf{e}, Hall conductivity ($\sigma_{xy}$) along the c-axis (in-plane) and b-axis (out-of-plane) at varying pressures ($\leq$0.95 GPa). A significant anomalous Hall conductivity ($\sigma^A_{xy}$) develops with the applied magnetic field in both directions, persisting up to 8 T. This is attributed to Berry curvature effects caused by the redistribution of Weyl nodes under applied magnetic fields. The magnitude of $\sigma^A_{xy}$ diminishes with increasing pressure for both directions, indicating a suppression of Berry curvature. \textbf{f}, Extracted anomalous Hall conductivity ($\sigma^A_{xy}$) from (\textbf{d}, \textbf{e}) vs. polarization magnitude. \textbf{g}, \textbf{h}, Polarization strength $\left|\gamma'\right|$ (derived from nonreciprocal signals) vs. magnetic field at different pressures. \textit{Insets}: The corresponding polarization directions. \textbf{i}, Nonlinear Hall signals, plotted as $E_y^{2\omega}/E_x^2$ (proportional to Berry curvature dipole density) vs. polarization strength $\left|\gamma'\right|$, with the magnetic field applied along the b-axis.}\label{Fig. 3
  }
  \label{fig:fig3}
\end{figure}

\begin{figure}[htbp]
  \centering
  \includegraphics[width=1\linewidth]{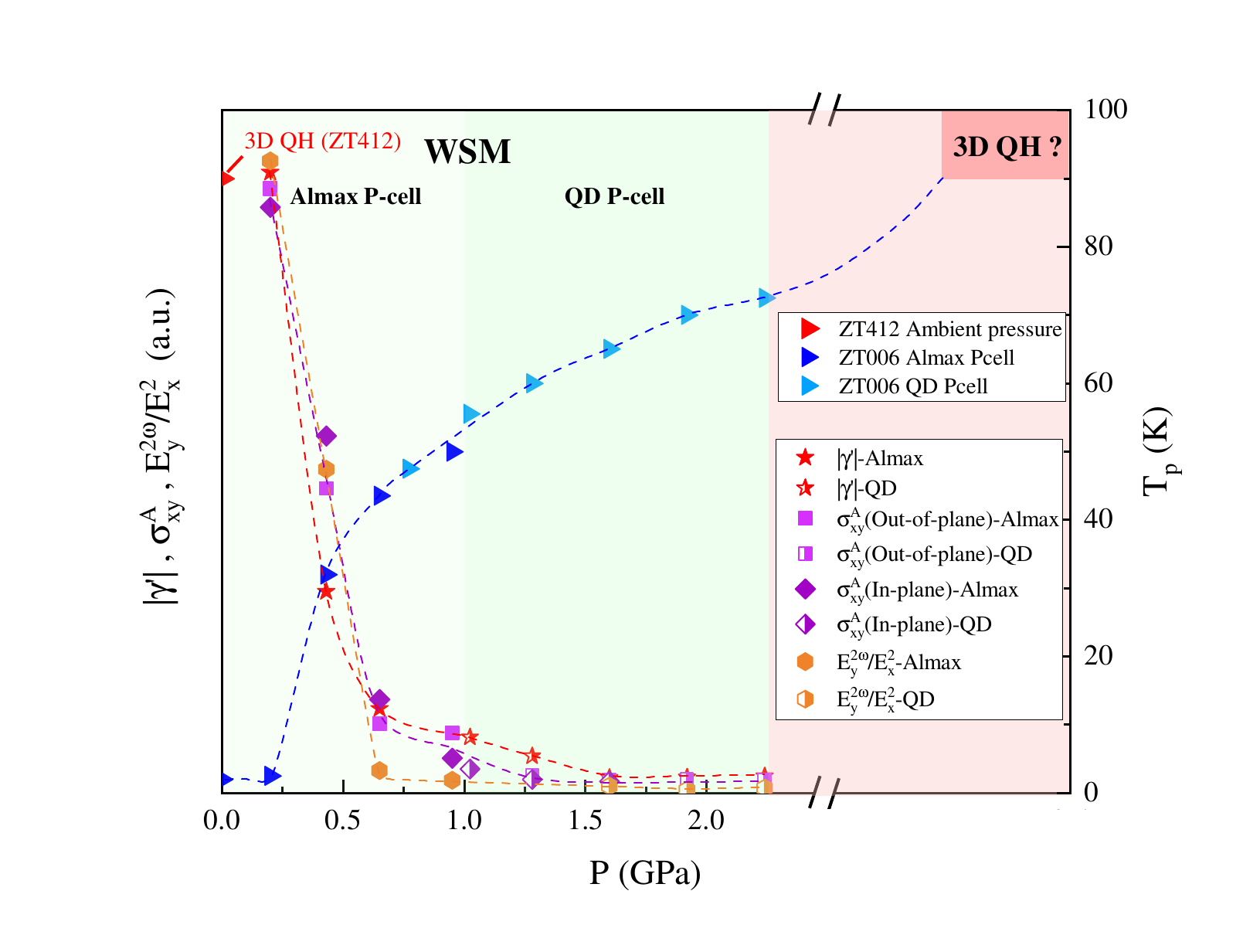}
  \caption{\textbf{Phase diagram of ZrTe$_5$.} 
 Increasing pressure suppresses key features of the WSM phase—including Berry curvature ($\sigma^A_{xy}$), Berry curvature dipole density ($E_y^{2\omega}/E_x^2$), and polarization magnitude ($\left|\gamma'\right|$). Simultaneously, the rise in $\text{\rmfamily T}_{\text{\rmfamily p}}$ (blue dashed line) signals a progressive transition into the 3D QH phase. To correct for a systematic error, data from the QD pressure cell (except for the peak temperature, \textit{$\text{\rmfamily T}_{\text{\rmfamily p}}$}) was scaled by a factor to align it with the data from the Almax pressure cell.}\label{Fig. 4
  }
  \label{fig:fig4}
\end{figure}
\end{document}